\documentclass[conference]{IEEEtran}
\IEEEoverridecommandlockouts

\usepackage{cite}
\usepackage{amsmath,amssymb,amsfonts}
\usepackage{array}
\usepackage{graphicx}
\usepackage{textcomp}
\usepackage{xcolor}
\usepackage{multirow}
\usepackage{makecell}
\usepackage{hhline}
\usepackage[hidelinks]{hyperref}
\usepackage{orcidlink}

\begin{document}

\title{Enhancing Vehicular Network Performance Through Integrated RSU and UAV Deployment}

\author{
\IEEEauthorblockN{
Muhammad Ismail~\orcidlink{0009-0002-7268-5486}
}
\IEEEauthorblockA{
\textit{Department of Electrical Engineering}\\
\textit{University of Engineering and Technology (UET) Mardan}\\
Mardan 23200, Pakistan\\
m.ismail012018@gmail.com
}
\and
\IEEEauthorblockN{
Syed Luqman Shah~\orcidlink{0000-0001-5792-0842}
}
\IEEEauthorblockA{
\textit{Telecommunication and Networking (TeleCoN) Research Lab}\\
\textit{GIK Institute of Engineering Sciences and Technology}\\
Topi 23640, Pakistan\\
sayedluqmans@gmail.com
}
\and
\IEEEauthorblockN{
Fazal Muhammad~\orcidlink{0000-0003-0405-0083}
}
\IEEEauthorblockA{
\textit{Department of Electrical Engineering}\\
\textit{University of Engineering and Technology (UET) Mardan}\\
Mardan 23200, Pakistan\\
fazal.muhammad@uetmardan.edu.pk
}
\and
\IEEEauthorblockN{
Zeeshan Shafiq~\orcidlink{0000-0002-6888-5111}
}
\IEEEauthorblockA{
\textit{National Center of Artificial Intelligence (NCAI)}\\
\textit{University of Engineering and Technology (UET) Peshawar}\\
Peshawar 25000, Pakistan\\
zeeshanshafiq@uetpeshawar.edu.pk
}
}

\maketitle

\begin{abstract}
The increasing density of connected vehicles can place substantial pressure on fixed roadside infrastructure, particularly when the available communication resources become insufficient to accommodate temporary traffic surges. This paper investigates auxiliary unmanned aerial vehicle (UAV) assistance as a flexible mechanism for improving service availability and throughput in vehicular networks. Two representative network configurations are considered. In the first, an auxiliary UAV (UAVa) supplements two fixed roadside units (RSUs), whereas in the second, UAVa assists a heterogeneous infrastructure comprising one RSU and one UAV. Vehicle service is determined according to node coverage, the line-of-sight (LoS) probability of aerial links, and a prescribed signal-to-interference-plus-noise ratio (SINR) requirement. The resulting framework enables UAVa to accommodate eligible vehicles that cannot be adequately served by the primary infrastructure as the network load increases. Simulation results show that, under the considered configurations, the aerial nodes benefit from more favorable propagation conditions and achieve higher throughput than the fixed terrestrial RSU. Moreover, the introduction of UAVa increases the available service capacity under high vehicular loads, both for a purely terrestrial baseline and for a network already supported by an aerial node. These results demonstrate the potential of auxiliary UAV assistance as a flexible load-relief mechanism for capacity-constrained vehicular networks.
\end{abstract}


\section{Introduction}
The increasing expansion of smart transportation and the growing need for reliable and high-quality communication services pose significant problems in guaranteeing seamless connectivity, effective data transfer, and interference minimization~\cite{001}. Traditional vehicular networks rely mainly on stationary traditional Road-Side-Units (RSUs) as terrestrial Base Stations (BSs), which can result in inferior network performance in the face of changing traffic circumstances and user densities. Furthermore, the growing need for bandwidth-intensive applications and ultra-low latency necessitates novel approaches to improving vehicular network throughput and Quality of Service (QoS)~\cite{002, 003}. To address these issues and further enhance vehicular networks, an intriguing alternative appears: the integration of unmanned aerial vehicles (UAVs) as aerial base stations to supplement existing RSUs~\cite{Luqman, 004}. UAV-assisted networks provide a chance for line of sight (LoS) link establishment, dynamic deployment, higher mobility, on-demand coverage provisioning, reliable connectivity, scalability, flexibility, and adaptability to high-data-traffic locations or emergency scenarios~\cite{Luqman}. Strategic placement of RSUs and UAVs, combined with advanced algorithms and optimization techniques that consider traffic density, mobility, communication range, and QoS requirements, can significantly improve network efficiency, reduce communication delays, and mitigate interference. This strategy ushers in an intelligent and adaptive deployment scheme powered by real-time traffic patterns and machine learning, artificial intelligence, and optimization algorithms, ushering in a transformative era for vehicular networks~\cite{006}.\\
With the rapid development of intelligent transportation systems and the growing need for connected vehicles, the efficient deployment of RSUs and UAVs as airborne BS has become critical to ensuring smooth communication, improved QoS, and reduced interference. One of the primary benefits of RSU and UAV deployment is a large increase in network throughput performance. By strategically deploying RSUs and a UAVa in their respective dwelling cells, network resources can be distributed more evenly and effectively, resulting in lower latency and higher data transmission rates. Furthermore, the UAVa 's dynamic mobility enables it to supplement fixed-position RSUs by opportunistically assisting areas with high traffic density and increasing the overall network coverage~\cite{011,012}. By intelligently deploying these UAVs and RSUs, the network may better accommodate the various demands of different vehicle applications, including real-time traffic updates, safety-critical communications, and multimedia streaming~\cite{013}.\\
Traditional approaches rely heavily on heuristic-driven algorithms such as genetic algorithms and particle swarm optimization to obtain estimated answers. For example,~\cite{018} proposed a genetic algorithm-based RSU placement technique. However, scaling issues frequently plague these systems, resulting in placements that need more optimality. Data-driven approaches that use real-world vehicle traffic patterns and simulations are a modern countermeasure to these restrictions. ~\cite{019} developed a vehicle movement pattern-based data-driven RSU deployment algorithm. Although these data-driven paradigms show promise, they may need to capture vehicular networks' dynamic nature properly. Because of its potential to improve coverage and network performance, incorporating UAVs as airborne base stations has gained significant popularity within vehicle networks~\cite{020}. Several studies have begun investigating the deployment of UAV-assisted networks, aiming to extend communication range and assist mobile users.In ~\cite{021} proposed a framework for UAV-assisted vehicular communication to improve connectivity in metropolitan areas. However, many issues still need to be addressed, including the proper placement of UAVs to guarantee seamless handovers and efficient resource allocation~\cite{022}. Prevalent techniques frequently lean towards static UAV placements, oblivious to the dynamic requirements of networks and the possible stumbling blocks provided by interference concerns. Interference management is a critical challenge in vehicular networks, particularly when UAVs and RSUs are involved~\cite{023}. Prior research efforts have painstakingly investigated a variety of interference mitigation approaches to improve network performance. Authors in~\cite{024} have developed a power control system to reduce interference in UAV-assisted vehicular networks. While such solutions demonstrate gains, they may need to account for numerous RSUs and UAVs, resulting in inefficient resource allocation and limited network throughput increases. This paper presents an integrated framework for the optimal deployment of RSUs and UAVa in vehicular networks to solve the inadequacies of previous techniques. The method employs a hybrid optimization strategy that blends heuristic-based algorithms with data-driven insights from real-world traffic patterns. The suggested strategy seeks to improve coverage and connectivity while lowering infrastructure costs by optimizing RSU placement~\cite{026}. The authors in~\cite{ismail01} introduce a novel end-to-end latency model for 5G Vehicle-to-Network (V2N) and Vehicle-to-Network-to-Vehicle (V2N2V) communications, assessing the impact of various 5G network deployments and configurations on latency for advanced V2X services, revealing the challenges and benefits of different deployment strategies. The authors in~\cite{ismail02} propose an analytical model to assess the potential of 5G's flexible New Radio interface for meeting stringent latency requirements of advanced V2X services, employing V2N2V communications, while accounting for various radio configurations and scenarios.\\
However, in very congested environments and emergency situations, the integration of UAV-assistant (UAVa) into the existing vehicular network infrastructure and their resource management is challenging. UAVa is dynamically deployed at some higher altitude in emergency situation to assist the existing network in their resources. This results in a more reliable and effective vehicular communication system, providing a smooth ride for both drivers and passengers. Furthermore, the best deployment of UAVa reduces the likelihood of packet loss and communication disruptions, resulting in a more stable and trustworthy vehicle network~\cite{014}.\\
\subsection{Novelty and Contributions}
Our main contributions of this work are summarized as follows:
\begin{itemize}
    \item We consider two complementary UAV-assisted vehicular-network configurations: Scenario~I, in which UAVa assists two fixed RSUs, and Scenario~II, in which UAVa assists a heterogeneous infrastructure consisting of one RSU and one UAV.
    \item We formulate a coverage- and channel-aware vehicle-association procedure based on infrastructure coverage, UAV--vehicle LoS probability, and an SINR decoding threshold. The auxiliary UAV is used to support vehicles that cannot be adequately served by the primary infrastructure under high-load conditions.
    \item We evaluate the throughput of the individual infrastructure nodes and the aggregate network under varying vehicle loads, thereby quantifying the benefit of auxiliary aerial assistance in the two considered scenarios.
\end{itemize}

The remainder of this paper is organized as follows. Section~\ref{Sec3:SystemModel} introduces the system model and the two considered network scenarios. Section~\ref{Sec4:ProposedScheme} describes the proposed association and transmission procedure. Section~\ref{Sec5:SimulationResults} presents the simulation setup and discusses the numerical results. Finally, Section~\ref{Sec6:Conclusion} concludes the paper and outlines possible directions for future work.

\section{System Model}
\label{Sec3:SystemModel}

We consider a time-slotted vehicular network in which vehicles are served by a combination of fixed and aerial communication infrastructure. At time slot $t$, let $\mathcal{V}^{t}$ denote the set of active vehicles, with $v\in\mathcal{V}^{t}$ denoting an arbitrary vehicle. Vehicle arrivals are modeled according to a Poisson process with mean arrival rate $\lambda_{\mathrm{v}}$. The available infrastructure has finite communication resources and, therefore, may not be able to simultaneously serve all vehicles when the traffic load becomes sufficiently large.

To provide additional service under such conditions, we introduce an auxiliary UAV, denoted by UAVa. In contrast to the primary infrastructure, UAVa is used as an on-demand aerial node and assists vehicles that cannot be adequately served by the initially deployed nodes. This results in the two network configurations considered in this work. In Scenario~I, UAVa supplements two fixed RSUs, whereas in Scenario~II it supplements a heterogeneous infrastructure consisting of one RSU and one UAV. The two scenarios are described next.

\begin{figure}[t]
\centering
\includegraphics[width=\linewidth]{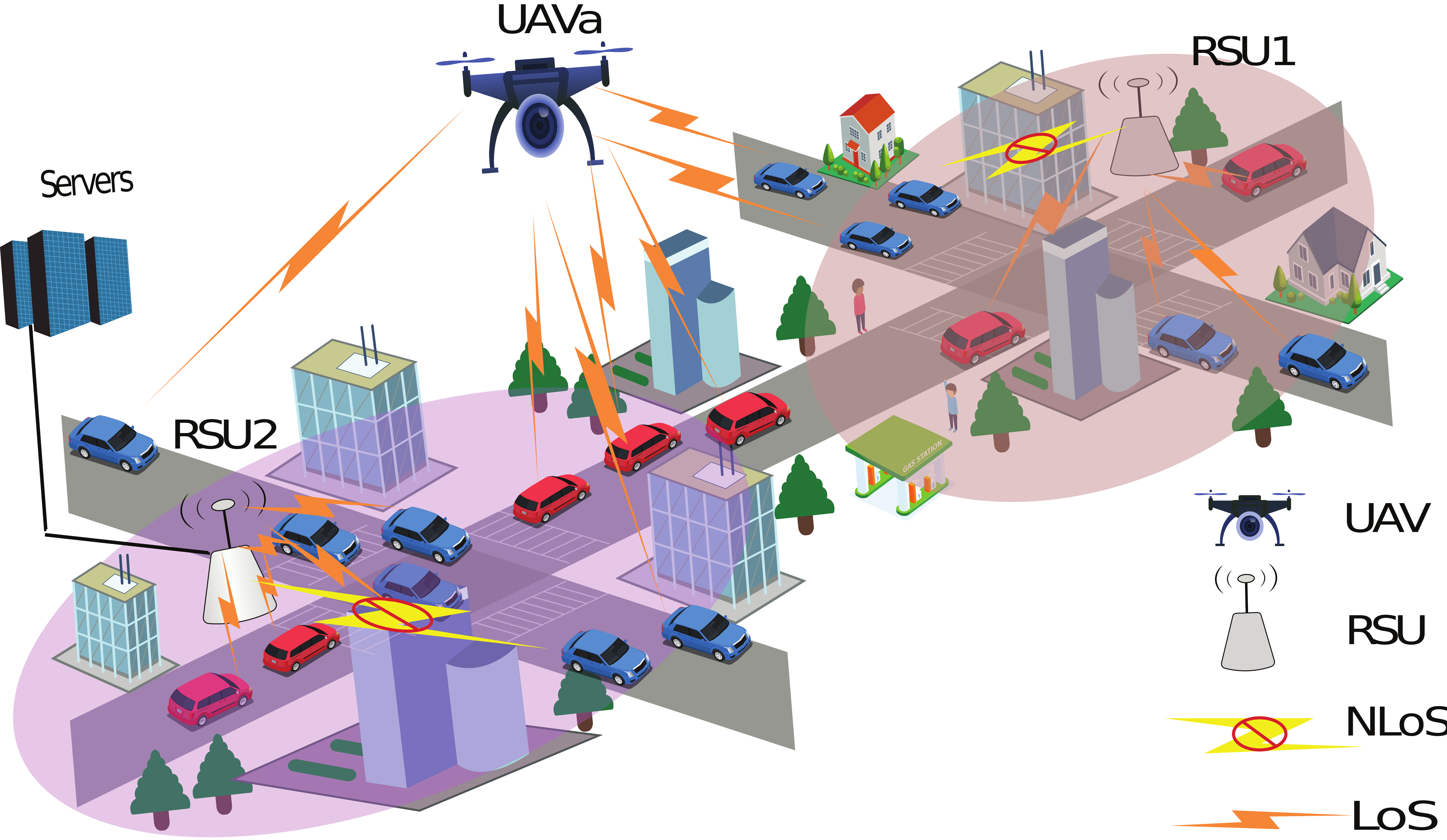}
\caption{System model for Scenario~I with RSU1, RSU2, and the auxiliary UAV (UAVa).}
\label{Fig1: System Model Scenario-I}
\end{figure}

\subsection{Scenario I}
\label{System Scenario I}

Scenario~I consists of two fixed infrastructure nodes, RSU1 and RSU2, together with UAVa, as illustrated in Fig.~\ref{Fig1: System Model Scenario-I}. Under nominal traffic conditions, RSU1 and RSU2 provide service to vehicles located within their respective coverage regions. Since both RSUs have finite communication resources, however, increasing vehicle density may create a service demand that exceeds the available terrestrial resources.

In such a case, UAVa provides an additional aerial service opportunity for vehicles that cannot be accommodated by the two RSUs. Owing to its elevated position and deployment flexibility, UAVa can complement the fixed terrestrial infrastructure and redistribute part of the traffic load during periods of congestion. Hence, Scenario~I represents a conventional RSU-based vehicular network augmented by an auxiliary UAV when additional communication resources are required.

\begin{figure}[t]
\centering
\includegraphics[width=\linewidth]{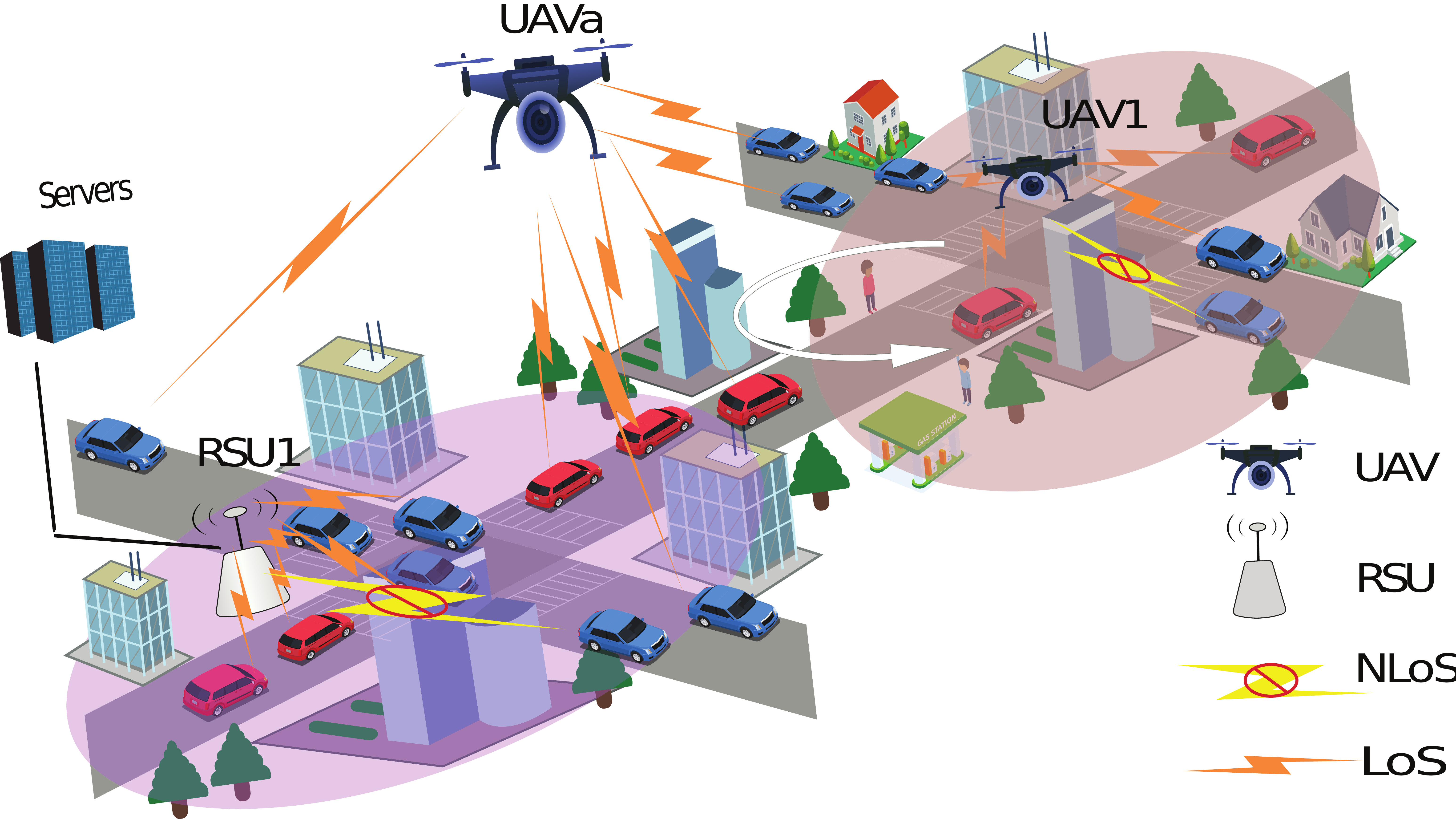}
\caption{System model for Scenario~II with RSU1, UAV1, and the auxiliary UAV (UAVa).}
\label{Fig1: System Model Scenario-II}
\end{figure}

\subsection{Scenario II}
\label{System Scenario II}

Scenario~II considers a heterogeneous primary infrastructure consisting of RSU1 and UAV1, together with UAVa, as shown in Fig.~\ref{Fig1: System Model Scenario-II}. Unlike Scenario~I, one of the primary serving nodes is already aerial. Thus, RSU1 provides terrestrial connectivity, while UAV1 provides an additional aerial coverage opportunity under normal network operation.

As the traffic load increases, the combined resources of RSU1 and UAV1 may become insufficient to serve all eligible vehicles. UAVa is then employed as an auxiliary aerial node to accommodate part of the excess demand. This scenario therefore examines whether on-demand UAV assistance remains beneficial when the underlying vehicular network already includes an aerial serving node. Comparing Scenarios~I and~II allows us to assess the contribution of UAVa under both purely terrestrial and heterogeneous terrestrial--aerial baseline infrastructures.

\section{Proposed Scheme}
\label{Sec4:ProposedScheme}

\begin{figure}[t]
    \centering
    \includegraphics[width=\linewidth]{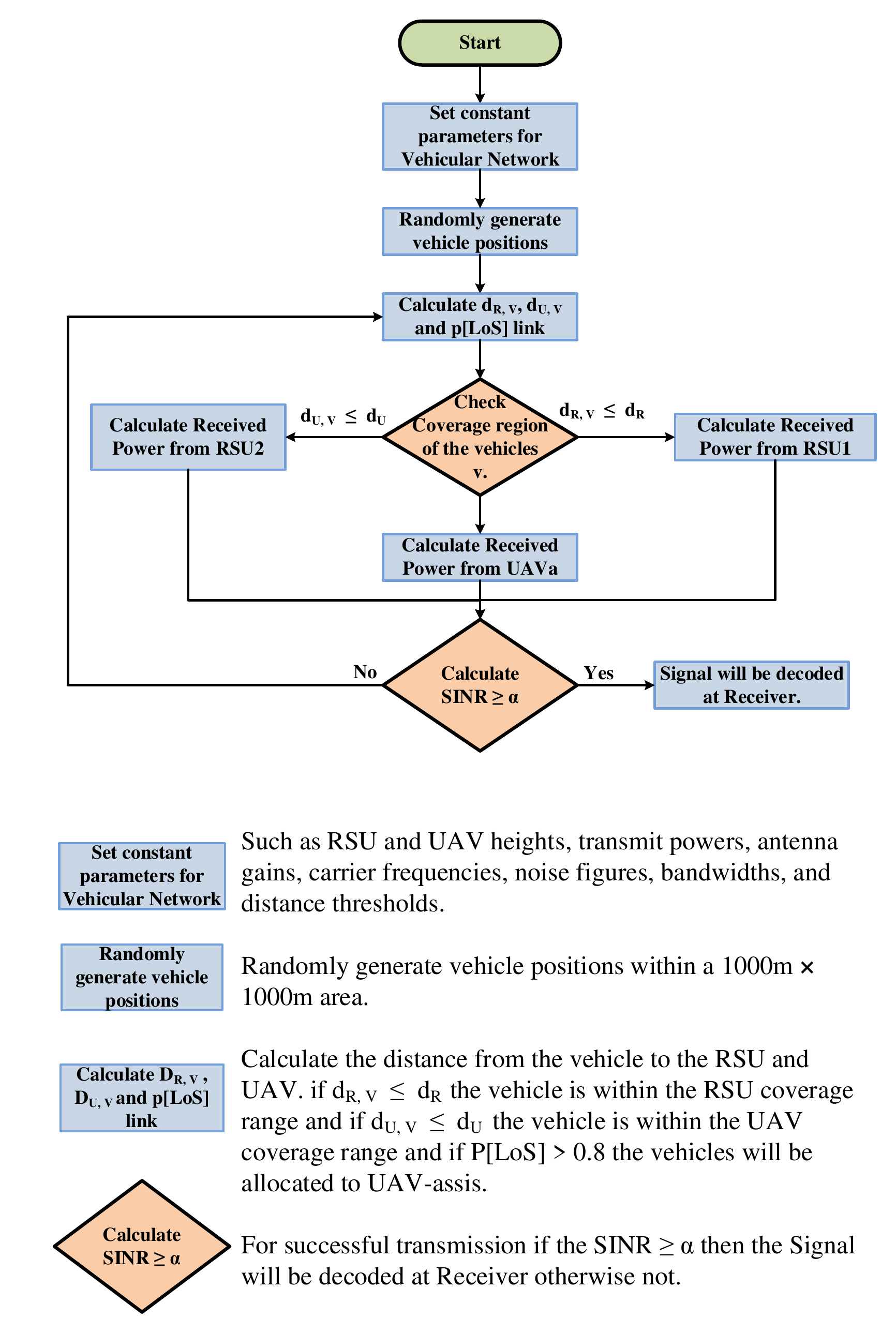}
    \caption{Service and channel-allocation procedure for Scenario~I.}
    \label{Fig3:FlowChartScenarioI}
\end{figure}

\begin{figure}[t]
    \centering
    \includegraphics[width=\linewidth]{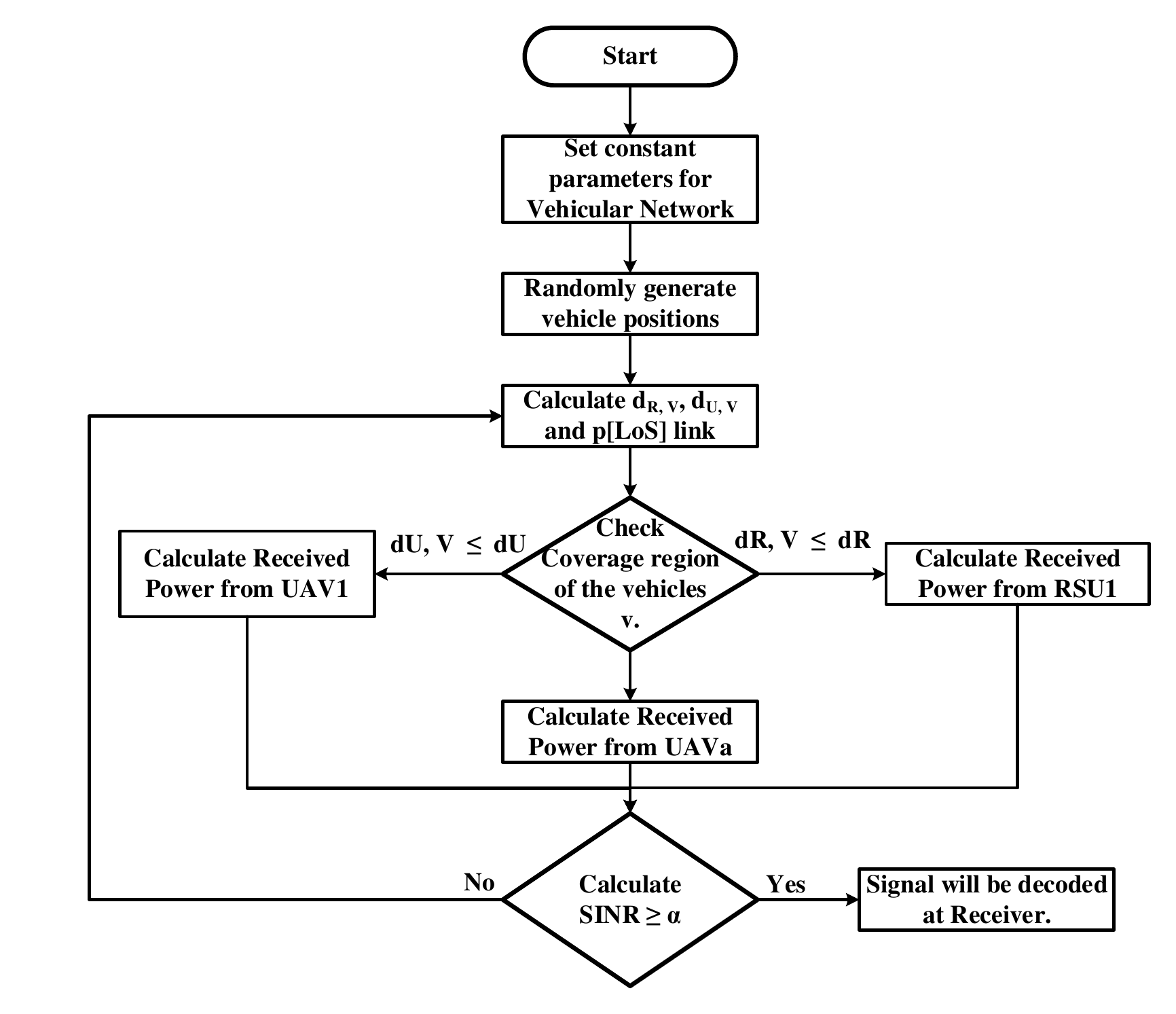}
    \caption{Service and channel-allocation procedure for Scenario~II.}
    \label{Fig3:FlowChartScenarioII}
\end{figure}

We now describe the service and channel-allocation procedure used in the two network configurations introduced in Section~\ref{Sec3:SystemModel}. The procedure is illustrated in Figs.~\ref{Fig3:FlowChartScenarioI} and~\ref{Fig3:FlowChartScenarioII}. The two scenarios follow the same association principle and differ only in the set of primary serving nodes. In Scenario~I, the primary infrastructure consists of RSU1 and RSU2, whereas in Scenario~II it consists of RSU1 and UAV1. In both cases, UAVa is invoked when the primary infrastructure cannot adequately serve an eligible vehicle.

For compactness, let $\mathcal{S}_{1}=\{\mathrm{RSU1},\mathrm{RSU2},\mathrm{UAVa}\}$ and $\mathcal{S}_{2}=\{\mathrm{RSU1},\mathrm{UAV1},\mathrm{UAVa}\}$ denote the candidate serving-node sets in Scenarios~I and~II, respectively. A node $\ell\in\mathcal{S}_{i}$ can serve vehicle $v$ at time slot $t$ only if the vehicle lies within its coverage region, i.e.,
\begin{equation}
    d_{\ell,v}^{t}\leq r_{\ell},
    \label{eq:coverage}
\end{equation}
where $d_{\ell,v}^{t}$ is the distance between node $\ell$ and vehicle $v$, and $r_{\ell}$ is the coverage radius of node $\ell$. Thus, the coverage test first eliminates serving nodes that are geographically unable to support the considered vehicle. Among the remaining candidates, channel admissibility is determined from the propagation conditions and the resulting SINR, as described next.

\subsection{Air-to-Ground Link Model}

The elevated position of a UAV generally provides a larger probability of establishing an LoS link than a terrestrial serving node. For a UAV $U\in\{\mathrm{UAV1},\mathrm{UAVa}\}$ and vehicle $v$, we model the LoS probability at time slot $t$ as
\begin{equation}
    p_{\mathrm{LoS},U,v}^{t}
    =
    \frac{1}
    {1+\varepsilon_{1}
    \exp\!\left[
    -\varepsilon_{2}
    \left(
    \theta_{U,v}^{t}-\varepsilon_{1}
    \right)
    \right]},
    \label{eq:los_probability}
\end{equation}
where $\varepsilon_{1}$ and $\varepsilon_{2}$ are environment-dependent parameters and $\theta_{U,v}^{t}$ denotes the elevation angle from vehicle $v$ to UAV $U$. In particular, if $\rho_{U,v}^{t}$ denotes the horizontal separation between the UAV and the vehicle, while $h_U$ and $h_v$ denote their respective heights, then
\begin{equation}
    \theta_{U,v}^{t}
    =
    \frac{180}{\pi}
    \tan^{-1}
    \left(
    \frac{h_U-h_v}
    {\rho_{U,v}^{t}}
    \right).
    \label{eq:elevation_angle}
\end{equation}
Hence, increasing the elevation angle generally increases the probability of an LoS connection. In the proposed scheme, an aerial link is regarded as sufficiently favorable when
\begin{equation}
    p_{\mathrm{LoS},U,v}^{t}\geq p_{\mathrm{th}},
    \label{eq:los_threshold}
\end{equation}
where $p_{\mathrm{th}}\in[0,1]$ is the prescribed minimum LoS-probability threshold.

To distinguish the propagation characteristics of terrestrial and aerial links, we model their large-scale channel power gains separately. For a terrestrial RSU $R$, the gain of the RSU--vehicle link is written as
\begin{equation}
    G_{R,v}^{t}
    =
    \beta_{R}
    \left(d_{R,v}^{t}\right)^{-\eta_{R}},
    \label{eq:rsu_channel_gain}
\end{equation}
where $\eta_{R}>0$ is the ground-to-ground path-loss exponent and $\beta_{R}>0$ collects the corresponding reference-distance gain and other large-scale attenuation factors.

For an aerial link, let $S_{U,v}^{t}\in\{\mathrm{L},\mathrm{N}\}$ denote its propagation state, where $\mathrm{L}$ and $\mathrm{N}$ represent LoS and non-LoS (NLoS), respectively. The probability of the LoS state is given by~\eqref{eq:los_probability}. Conditioned on the propagation state, the UAV--vehicle channel power gain is modeled as
\begin{equation}
    G_{U,v}^{t}
    =
    \begin{cases}
        \beta_{\mathrm{L}}
        \left(d_{U,v}^{t}\right)^{-\eta_{\mathrm{L}}},
        & S_{U,v}^{t}=\mathrm{L},
        \\[1mm]
        \beta_{\mathrm{N}}
        \left(d_{U,v}^{t}\right)^{-\eta_{\mathrm{N}}},
        & S_{U,v}^{t}=\mathrm{N},
    \end{cases}
    \label{eq:uav_channel_gain}
\end{equation}
where $\eta_{\mathrm{L}}$ and $\eta_{\mathrm{N}}$ denote the LoS and NLoS path-loss exponents, respectively, and $\beta_{\mathrm{L}}$ and $\beta_{\mathrm{N}}$ denote the corresponding reference-distance power gains. Typically, the NLoS state incurs stronger attenuation than the LoS state, which is reflected by the corresponding channel parameters.

\subsection{Channel Selection and SINR Constraint}

Let $\mathcal{C}=\{1,\ldots,N_{\mathrm{ch}}\}$ denote the set of orthogonal channels available to the network, where $N_{\mathrm{ch}}$ is the total number of channels. For candidate serving node $\ell$ and vehicle $v$, the received SINR on channel $c\in\mathcal{C}$ at time slot $t$ is
\begin{equation}
    \gamma_{\ell,v,c}^{t}
    =
    \frac{
        P_{\ell}G_{\ell,v}^{t}
    }{
        \sigma^{2}
        +
        I_{\ell,v,c}^{t}
    },
    \label{eq:sinr}
\end{equation}
where $P_{\ell}$ is the transmit power of node $\ell$, $G_{\ell,v}^{t}$ is the corresponding large-scale channel power gain, $\sigma^{2}$ is the receiver-noise power over the considered bandwidth, and $I_{\ell,v,c}^{t}$ is the aggregate co-channel interference experienced by vehicle $v$ on channel $c$.

More explicitly, the interference term accounts only for simultaneously active transmitters using the same channel and can be expressed as
\begin{equation}
    I_{\ell,v,c}^{t}
    =
    \sum_{\substack{j\in\mathcal{S}_{i}\\j\neq\ell}}
    a_{j,c}^{t}
    P_{j}G_{j,v}^{t},
    \label{eq:interference}
\end{equation}
where $a_{j,c}^{t}\in\{0,1\}$ is an activity indicator that equals one when node $j$ transmits on channel $c$ during time slot $t$, and zero otherwise. Consequently, transmissions over different orthogonal channels do not contribute to the co-channel interference term.

A link is considered admissible only when its instantaneous SINR satisfies
\begin{equation}
    \gamma_{\ell,v,c}^{t}\geq\gamma_{\mathrm{th}},
    \label{eq:sinr_threshold}
\end{equation}
where $\gamma_{\mathrm{th}}$ is the minimum SINR required for successful reception. If the condition in~\eqref{eq:sinr_threshold} is not satisfied, the considered channel is rejected and the vehicle attempts to access another available channel according to the channel-access procedure shown in Figs.~\ref{Fig3:FlowChartScenarioI} and~\ref{Fig3:FlowChartScenarioII}. Thus, the SINR threshold is not used to ``remove'' interference; rather, it provides an explicit criterion for determining whether the resulting signal quality is sufficient for communication.

\subsection{Instantaneous Throughput}

Once a vehicle has been associated with a serving node and an admissible channel has been identified, the corresponding instantaneous link throughput is computed from its SINR. Let $B$ denote the bandwidth allocated to one channel. The instantaneous throughput from node $\ell$ to vehicle $v$ on channel $c$ is
\begin{equation}
    R_{\ell,v,c}^{t}
    =
    B\log_{2}
    \left(
    1+\gamma_{\ell,v,c}^{t}
    \right),
    \label{eq:instantaneous_throughput}
\end{equation}
measured in bit/s. Equation~\eqref{eq:instantaneous_throughput} therefore maps the instantaneous received signal quality into the corresponding Shannon rate over the allocated bandwidth. A vehicle contributes to the served throughput only when the coverage condition in~\eqref{eq:coverage}, the required aerial-link condition in~\eqref{eq:los_threshold} whenever applicable, and the SINR condition in~\eqref{eq:sinr_threshold} are simultaneously satisfied.

For a serving node $\ell$, let $\mathcal{V}_{\ell,c}^{t}$ denote the set of vehicles successfully served by that node on channel $c$ during time slot $t$. Its aggregate instantaneous throughput is then
\begin{equation}
    R_{\ell}^{t}
    =
    \sum_{c\in\mathcal{C}}
    \sum_{v\in\mathcal{V}_{\ell,c}^{t}}
    R_{\ell,v,c}^{t}.
    \label{eq:node_throughput}
\end{equation}
Accordingly, the network-wide throughput is obtained by summing~\eqref{eq:node_throughput} over all active serving nodes in the considered scenario. This quantity is used in Section~\ref{Sec5:SimulationResults} to compare the load-handling capability of the two network configurations and to quantify the additional service provided by UAVa.

\section{Simulation Results and Discussion}
\label{Sec5:SimulationResults}

We now evaluate the performance of the proposed scheme under the two network configurations introduced in Section~\ref{Sec3:SystemModel}. The simulations are implemented in MATLAB R2021b. The network load is varied by changing the number of active vehicles, while vehicle service is determined according to the coverage, LoS-probability, and SINR conditions described in Section~\ref{Sec4:ProposedScheme}. For each successfully served vehicle, the instantaneous throughput is computed according to~\eqref{eq:instantaneous_throughput}, and the corresponding node throughput is obtained from~\eqref{eq:node_throughput}.

The considered network employs $N_{\mathrm{ch}}=10$ orthogonal channels and a channel-access probability of $q=0.2$. The minimum SINR required for successful reception is set to $\gamma_{\mathrm{th}}=10$~dB. For aerial links, the minimum admissible LoS probability is set to $p_{\mathrm{th}}=0.8$, consistently with~\eqref{eq:los_threshold}. The locations of the serving nodes and the remaining simulation parameters are summarized in Table~\ref{tab:simulation_parameters}.

\begin{table}[t]
\centering
\caption{Simulation parameters.}
\label{tab:simulation_parameters}
\renewcommand{\arraystretch}{1.1}
\begin{tabular}{|l|c|}
\hline
\textbf{Parameter} & \textbf{Value} \\
\hline\hline
RSU1 coordinates $(x,y,z)$ [m] & $(125,-5,30)$ \\
\hline
RSU2 coordinates $(x,y,z)$ [m] & $(125,10,30)$ \\
\hline
UAV1 coordinates $(x,y,z)$ [m] & $(125,10,30)$ \\
\hline
UAVa coordinates $(x,y,z)$ [m] & $(250,0,40)$ \\
\hline
Number of orthogonal channels, $N_{\mathrm{ch}}$ & $10$ \\
\hline
Channel-access probability, $q$ & $0.2$ \\
\hline
SINR threshold, $\gamma_{\mathrm{th}}$ & $10$~dB \\
\hline
LoS-probability threshold, $p_{\mathrm{th}}$ & $0.8$ \\
\hline
\end{tabular}
\end{table}

\begin{figure}[t]
\centering
\includegraphics[width=\linewidth]{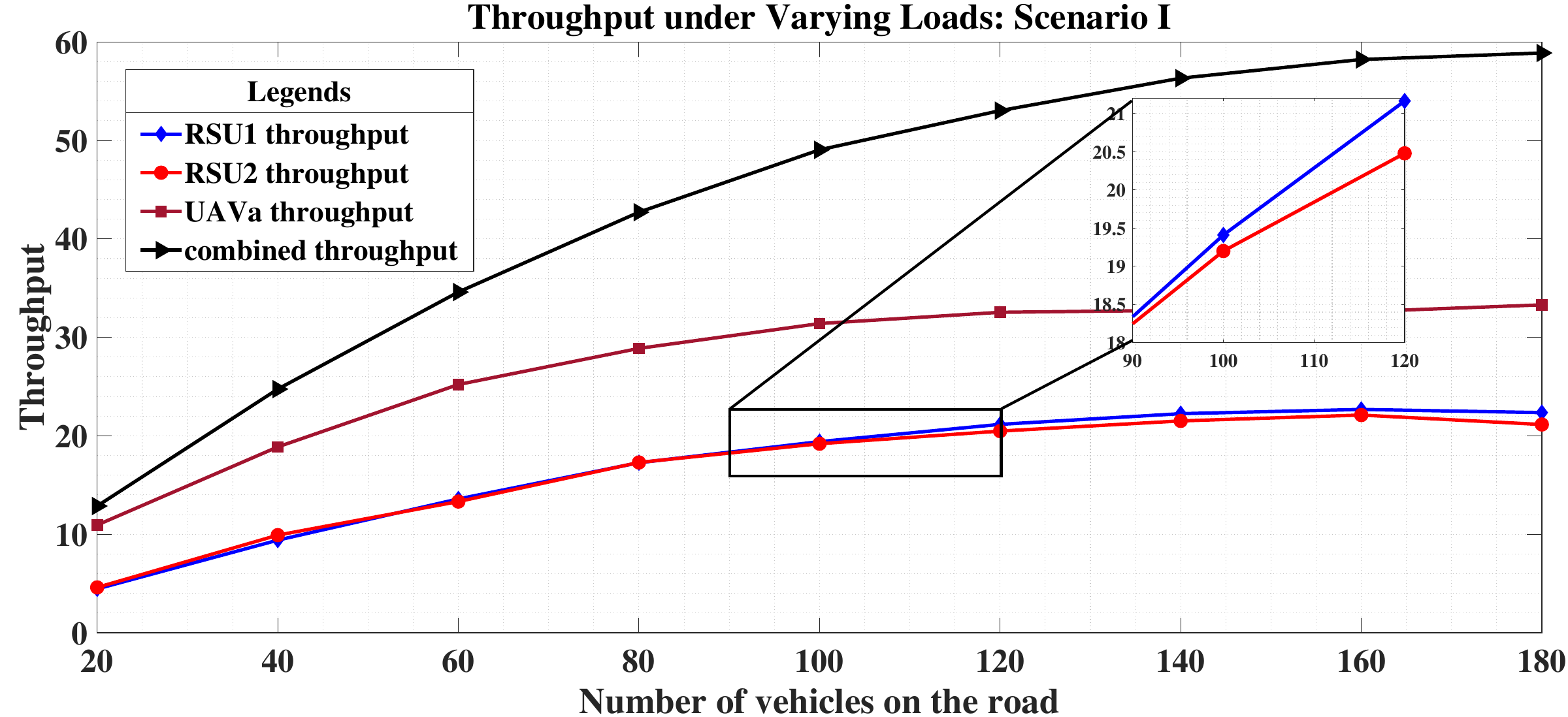}
\caption{Throughput performance versus network load for Scenario~I.}
\label{fig:scenario_I}
\end{figure}

\subsection{Scenario I}

Fig.~\ref{fig:scenario_I} shows the throughput achieved by RSU1, RSU2, and UAVa as the network load increases. The two RSUs exhibit similar throughput behavior, which is consistent with their comparable role as fixed terrestrial serving nodes and with the common association and channel-access rules adopted in the proposed scheme. Their throughput becomes increasingly constrained as the number of active vehicles grows, since each RSU can serve only the vehicles satisfying the corresponding coverage and link-quality requirements.

In contrast, UAVa achieves a higher throughput over the considered load range. This behavior can be explained by the more favorable propagation geometry provided by its elevated position. In particular, a larger fraction of candidate UAVa--vehicle links can satisfy the LoS requirement $p_{\mathrm{LoS},U,v}^{t}\geq p_{\mathrm{th}}$, thereby creating additional service opportunities for vehicles that cannot be accommodated by the fixed RSUs. The benefit becomes more pronounced as the traffic load increases, since UAVa progressively absorbs part of the demand that would otherwise remain unsupported by the terrestrial infrastructure.

The important observation is therefore not merely that UAVa attains a larger node throughput, but that it provides an additional degree of freedom for load redistribution. The fixed RSUs continue to provide the primary terrestrial service, while UAVa complements them when the available terrestrial resources become insufficient. This behavior is precisely the operating principle underlying Scenario~I.

\begin{figure}[t]
\centering
\includegraphics[width=\linewidth]{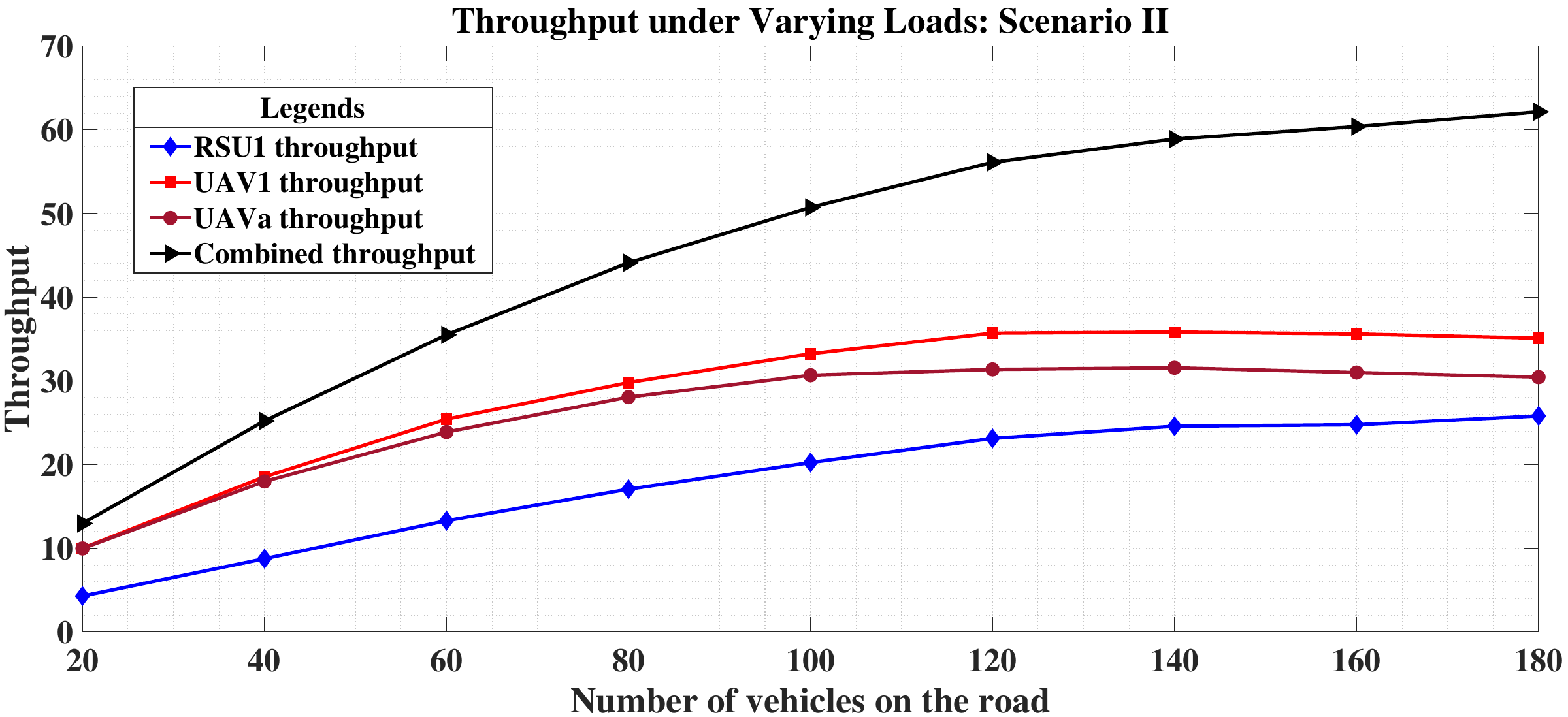}
\caption{Throughput performance versus network load for Scenario~II.}
\label{fig:scenario_II}
\end{figure}

\subsection{Scenario II}

We next consider Scenario~II, in which the primary infrastructure already contains an aerial serving node. Fig.~\ref{fig:scenario_II} reports the throughput of RSU1, UAV1, and UAVa, together with the resulting aggregate network throughput, as the offered load increases.

RSU1 exhibits the lowest throughput among the serving nodes in the considered configuration. Since RSU1 is fixed at the terrestrial level, its service opportunities are determined primarily by its coverage region and the quality of the corresponding ground-to-ground links. UAV1, on the other hand, benefits from an elevated position and therefore provides more favorable propagation conditions for a larger subset of vehicles. Consequently, UAV1 supports a larger traffic load than RSU1 under the considered network geometry.

The introduction of UAVa provides an additional gain beyond that already obtained from UAV1. When the service capability of RSU1 and UAV1 becomes insufficient, UAVa can accommodate additional vehicles whose coverage, LoS-probability, and SINR requirements are satisfied. As a result, the aggregate throughput increases with the contribution of all three serving nodes.

Comparing Figs.~\ref{fig:scenario_I} and~\ref{fig:scenario_II} reveals the main role of UAVa in the proposed architecture. In Scenario~I, UAVa supplements a purely terrestrial baseline, whereas in Scenario~II it supplements a network that already contains an aerial node. In both cases, its contribution is primarily associated with additional service availability under increasing network load. The results therefore indicate that auxiliary aerial assistance can remain beneficial even when UAV support is already part of the underlying vehicular infrastructure.

\section{Conclusion and Future Work}
\label{Sec6:Conclusion}

This paper investigated the use of an auxiliary UAV to complement the communication infrastructure of vehicular networks under increasing traffic load. Two representative configurations were considered: a terrestrial baseline comprising two RSUs and a heterogeneous baseline comprising one RSU and one UAV. In both cases, UAVa provides additional service when the primary infrastructure cannot adequately accommodate the active vehicles. The proposed procedure combines coverage constraints, the LoS characteristics of aerial links, and an SINR requirement to determine whether a vehicle can be successfully served.

The simulation results demonstrate the benefit of introducing an auxiliary aerial node under the considered network configurations. Owing to its favorable propagation geometry, UAVa provides additional service opportunities and increases the available network throughput as the vehicular load grows. The results further show that this benefit is not limited to a purely terrestrial baseline; auxiliary UAV assistance remains useful even when an aerial node is already part of the primary infrastructure. These observations highlight the potential of on-demand aerial assistance as a flexible mechanism for accommodating temporary increases in communication demand.

The present study considers fixed UAV locations and a predefined service procedure. A natural extension is therefore to jointly optimize UAV placement, trajectory, and resource allocation as functions of the spatial and temporal traffic distribution. Incorporating energy constraints, mobility-aware handover mechanisms, security considerations, and measurement-calibrated propagation models would further enable a more comprehensive assessment of the reliability, efficiency, and scalability of UAV-assisted vehicular networks.

\bibliographystyle{IEEEtran}
\bibliography{references.bib}

\end{document}